\documentclass[aps,prl,floatfix,superscriptaddress,onecolumn,footinbib]{revtex4-1}
\usepackage{graphicx}
\usepackage{dcolumn}
\usepackage{bm}
\usepackage{makecell}
\usepackage{bbold} 
\usepackage{color}
\usepackage{amsmath}
\usepackage{latexsym}
\usepackage{amssymb}
\usepackage{graphics,epstopdf}
\usepackage{hyperref}
\usepackage{xcolor}
\hypersetup{
    colorlinks = true,
    linkcolor =blue,
	citecolor=blue,
	urlcolor=blue
}

\newcommand{\ed}{\end{document}}
\newcommand{\beq}{\begin{equation}}
\newcommand{\eeq}{\end{equation}}

\usepackage{mathtools}
\usepackage{comment}

\begin{document}

\title{\fontsize{11.00pt}{8pt}\selectfont \textbf{Knot-Driven Spin Selectivity: Topological Chirality-Induced Robust Spin Polarization in Molecular Knots}}

\author{Xi Sun}
\affiliation{School of Physics and Wuhan National High Magnetic Field Center,
Huazhong University of Science and Technology, Wuhan 430074, People's Republic of China.}

\author{Kai-Yuan Zhang}
\affiliation{School of Physics and Wuhan National High Magnetic Field Center,
Huazhong University of Science and Technology, Wuhan 430074, People's Republic of China.}

\author{Shu-Zheng Zhou}
 \affiliation{School of Physics and Wuhan National High Magnetic Field Center,
Huazhong University of Science and Technology, Wuhan 430074, People's Republic of China.}

\author{Hua-Hua Fu}
\altaffiliation{Corresponding author.\\ hhfu@hust.edu.cn}
\affiliation{School of Physics and Wuhan National High Magnetic Field Center,
Huazhong University of Science and Technology, Wuhan 430074, People's Republic of China.}
\affiliation{Institute for Quantum Science and Engineering, Huazhong University of Science and Technology, Wuhan, Hubei 430074, China.}

\date{\today}

\begin{abstract}
\textbf{Compared to traditional structural chiral materials (e.g., DNA, helicene), topological chirality in trefoil knot molecules has demonstrated multiple remarkable advantages in chirality-induced spin selectivity (CISS), including ultra-high spin polarization of nearly 90\%, conductivity increased by two orders of magnitude, and high-temperature stability (up to 350°C). However, the underlying physical mechanism remains elusive. This work establishes, for the first time, a fundamental theoretical framework for topological chirality-induced spin selectivity (TCISS) in trefoil knot molecules and identifies the necessary conditions for knot-driven spin selectivity. Our calculation results reveal that a trefoil knot molecule can exhibit spin polarization exceeding 60\% along with significant conductivity. Notably, neither reducing the lattice number nor applying strain regulation significantly diminishes this ultra-high spin polarization, highlighting its robustness. Importantly, when the topological knot degenerates into a trivial structure, accompanied by the transition from topological chirality to structural chirality, the spin polarization sharply declines, demonstrating a strong correlation between the ultrahigh spin polarization and the knot topology. Our theory not only successfully elucidates the physical mechanism of TCISS, but also uncovers a new spin-polarized transport phenomenon termed knot-driven spin selectivity, offering new guiding principles for designing nonmagnetic materials for spintronics device applications.}

\end{abstract}
\maketitle

\newpage
\noindent
The interplay between the electron’s spin and the structural chirality gives rise to a novel spin-dependent transport phenomenon known as chirality-induced spin selectivity (CISS) {\color{blue}\cite{1,2,3,4,5,6}}. This effect was first observed in photoelectron spectroscopy experiments on DNA molecules {\color{blue}\cite{7}} and has since been verified in a variety of chiral materials, including helicenes {\color{blue}\cite{8,9,10}}, chiral crystals {\color{blue}\cite{11,12,13}}, and supramolecules {\color{blue}\cite{14,15,16,17,18,19}}. The most remarkable feature of CISS is the ability of chiral materials to produce spin-polarized charge transport with spin polarization as high as 80\% in the absence of an external magnetic field {\color{blue}\cite{20,21,22,23}}. Remarkably, the related spin-polarized state remains stable at high temperatures up to room temperature. This unique property offers opportunities for manipulating spin states and generating spin currents in non-magnetic materials, thereby establishing a new theoretical foundation for designing spintronic devices {\color{blue}\cite{22,23,24,25,26,27}}. Recent research has demonstrated the successful fabrication of room-temperature spin-emitting LEDs {\color{blue}\cite{28,29}} and high-efficiency photoelectric devices using chiral perovskites {\color{blue}\cite{30,31,32,33,34,35,36,37}} leveraging the CISS effect. Furthermore, CISS plays a crucial role in modulating chemical reactions {\color{blue}\cite{38,39,40}}, enhancing catalytic activity {\color{blue}\cite{41,42,43,44,45,46}}, and even influencing biological processes {\color{blue}\cite{47,48}}. To understand the underlying physical mechanisms of CISS, researchers have proposed numerous theoretical models {\color{blue}\cite{49,50,51,52,53,54,55,56,57,58,59,60,61,62,63,64}}. However, existing theoretical frameworks still exhibit systematic contradictions with or explanatory limitations when confronted by experimental observations {\color{blue}\cite{2}}, and a universal theory capable of comprehensively explaining the CISS effect in all chiral materials remains elusive {\color{blue}\cite{65}}.

Traditional CISS arises primarily from structural chirality, where helical configurations break spatial inversion symmetry, resulting in enantiomers with left- or right-handed chirality (Figure 1a). However, a recent experiment has revealed a novel manifestation of CISS in materials lacking a conventional helical chain molecule, i.e., molecular trefoil knots {\color{blue}\cite{66,67}} (Figure 1b). Despite they do not have classical stereogenic units, the entanglements of this kind of molecules generate a trefoil knot unit with topological chirality, dependent on the crossing arrangements (Figures 1c and 1d). This topological knot represents a higher-level spatial organization, ensuring their distinct chirality and significantly enhancing the spin selectivity stability. We therefore define this phenomenon as the topological chirality-induced spin selectivity (TCISS) effect. Compared to traditional CISS {\color{blue}\cite{1,2,3,4,5,6,7,8,9,10}}, TCISS observed in experiments demonstrates notable advantages: (i) ultra-high spin polarization of up to 90\%; (ii) higher conductivity by approximately 2 orders of magnitude; (iii) remarkable thermal stability, maintaining high spin polarization from room temperature to nearly 350°C {\color{blue}\cite{66}}. TCISS pioneers new pathways to understand the underlying physical mechanisms of CISS and develops a new spin-associated phenomenon, that is, \textbf{knot-driven spin selectivity}, marking a significant advance in the research topic of CISS. \textbf{However, a foundational theoretical framework for TCISS remains unestablished, and its microscopic physical mechanisms await elucidation.}

\begin{figure*}[t]%
\centering
\includegraphics[width=0.85\textwidth]{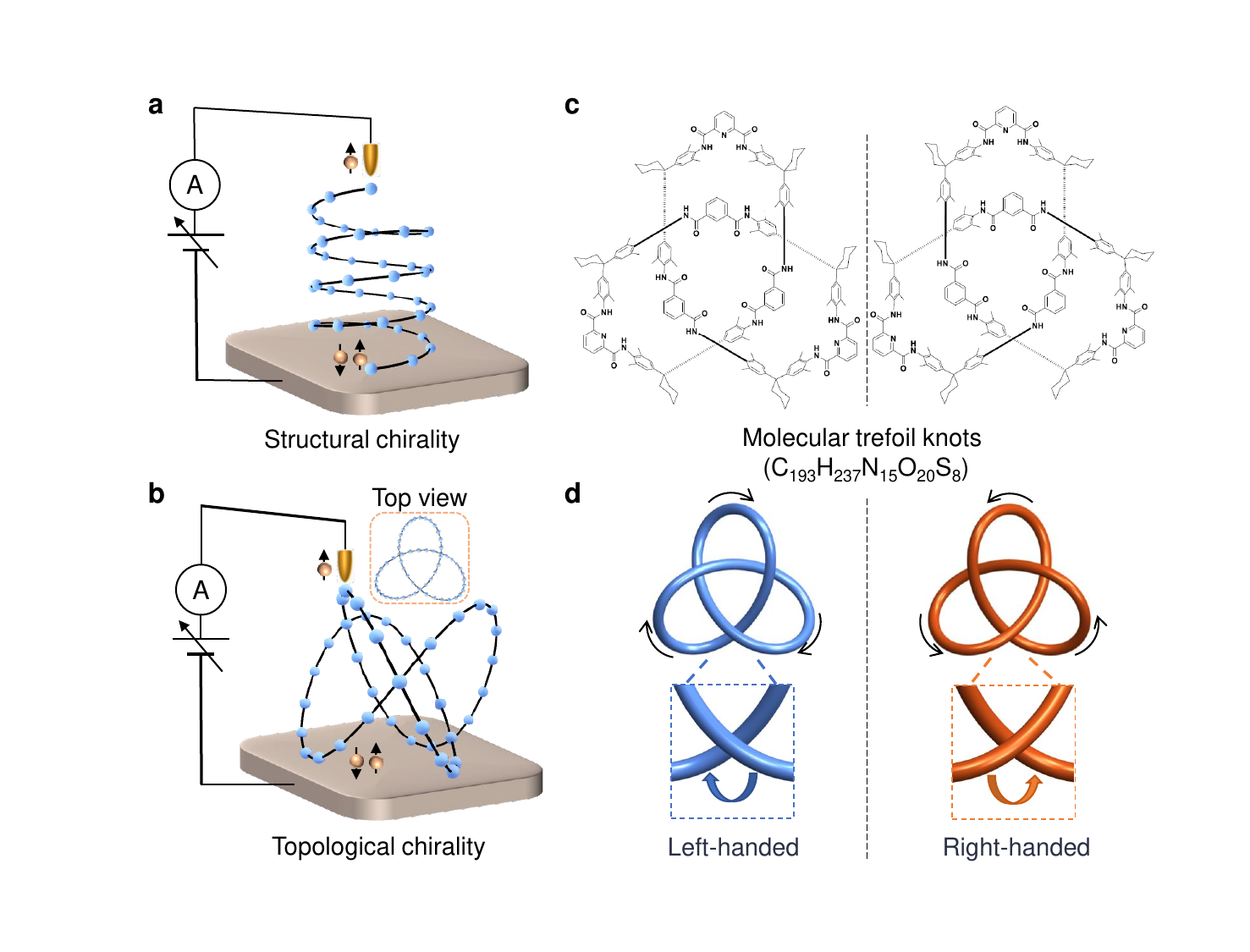}
\caption{\textbf{\textsf{Schematic descriptions of two different chiral materials to exhibit spin selectivity.}} \textsf{\textbf{a,} The conventional CISS device structured on a single helix chain with  structural chirality. \textbf{b,} The new-developed CISS device based on the trefoil knot molecules characterized by topological chirality. \textbf{c,} Paired enantiomers of real trefoil knot molecules possessing opposite chirality exampled by C$_{193}$H$_{237}$N$_{15}$O$_{20}$S$_{8}$. \textbf{d,} Trefoil knots with opposite topological chirality depending on the spatial arrangements of crossings.}}\label{fig1}
\end{figure*}

This work constructs a series of spin-dependent transport devices based on trefoil knot molecules to establish a foundational theory for the TCISS effect. To elucidate its underlying physics, the key challenge lies in clarifying the origin of spin polarization. Given the unique closed-loop topology of trefoil knot molecules, conventional CISS mechanisms, such as intramolecular spin-orbit coupling (SOC) {\color{blue}\cite{49,50,51,52}}, chiral-induced orbital polarization {\color{blue}\cite{58}}, spin-interface interactions {\color{blue}\cite{68,69,70}}, or non-Hermitian skin effects under open boundaries {\color{blue}\cite{64,71}}, are inapplicable. Considering the three-dimensionally curved transport channels within the trefoil knot {\color{blue}\cite{66,67}}, SOC originates from non-relativistic terms in the relativistic curved space theory. This SOC is intrinsically related to the geometry of the twisted transport paths and differs fundamentally from the conventional SOC applied in traditional CISS; therefore, we call it geometric SOC {\color{blue}\cite{72,73}}. Crucially, geometric SOC also depends on molecular chirality, and its strength varies spatially along the curved structure, explicitly mapping the topological knot geometry. In addition, geometric SOC values exceed those of conventional SOC by two orders of magnitude {\color{blue}\cite{72}}. These characteristics provide a physical mechanism for the ultra-high spin polarization high to 90\% observed in the TCISS measurements.\\

\noindent
{\fontsize{11pt}{10pt}\selectfont \textbf{Effective theoretical model with geometric SOC}}\\
\noindent
From a theoretical modeling perspective, the trefoil knot molecule can be simplified as a helical configuration exhibiting continuous spatial variation in curvature, forming the topological knot structure (Figure 1d). The twisted transport paths extending through 3D space result in enhanced spatial freedom for electron transmission, generating highly complex SOC characteristics. Crucially, conventional SOC explaining CISS, typically derived from the Dirac Lagrangian density in a flat spacetime electromagnetic field {\color{blue}\cite{49,50,51}}, is fundamentally inapplicable to these continuously deforming trefoil structures. Here, we demonstrate that the trefoil-knot arrangement of monomers generates geometric SOC. Specifically, within the helical structure of trefoil knot molecules, quantum interference fundamentally arises between direct ($i$-$j$) and indirect ($i$-$k$-$j$) paths for electron transmission, introducing an extra spin-dependent Berry phase during electron hopping (see Method). This phenomenon constitutes the microscopic origin of geometric SOC, with triad orientations governed by molecular chirality. Similarly to conventional SOC, geometric SOC plays a crucial role in inducing spin flipping during electron transport. To quantitatively characterize this mechanism, we discretize the trefoil knot molecules in a lattice-knot model: the closed helix comprises $N$ lattice sites with equal arc length (positions of lattice sites 1 and $N$ shown in Figure 2a; see Movie 1), where the spatial coordinates of the $i$th site are given by
\begin{equation}
\textbf{R}_i=\textbf{R}_i(\theta_i)=[r\mathrm{sin}{\theta_i}+R\mathrm{sin}{(2\theta_i)}, \zeta r\mathrm{cos}{\theta_i}-\zeta R\mathrm{cos}{(2\theta_i)}, h\mathrm{sin}(3\theta_i)]^T,
\end{equation}
where $R+r$ and $R-r$ denote the maximum and minimum radii, respectively, in the top view projection of the trefoil knot; $\zeta$ symbolizes the chirality of the molecular (left handed when $\zeta$=1 and right handed when $\zeta$=-1); $h$ represents the molecular height, and $\theta_i$ is the angle coordinate of the $i$th lattice. $\vec{T}_i$ and $\vec{N}_i$ are defined as the unit tangent and normal vectors of the twisted path (Supplemental Information (SI), Section A). Electrons traversing this trefoil knot experience dynamics analogous to a noninertial system with finite acceleration $\vec{a}_{i}=-\kappa c^2 \vec{N}_{i}$ (where $c$ is the speed of light), inducing the geometric SOC term $H_{\mathrm{geo}}=\hbar (\vec{a}\times\vec{p})\cdot \vec{\sigma}/{(4mc^{2})}=-\Sigma_{i} \hbar \kappa_{i} (\vec{N}_{i}\times\vec{p}_{i}) \cdot \vec{\sigma}/{(4m)}$ (where $\vec{p}_{i}$ is the electron momentum, $m$ is electron mass, and $\vec{\sigma}$ is the Pauli vector) in the equation. Crucially, this geometric SOC is explicitly governed by lattice-dependent curvature $\kappa_{i}$, which can be calculated by $\kappa_{i}=|\vec{R}'_{i} \times \vec{R}''_{i}|/|\vec{R}'_{i}|^{3}$. To validate this, we compute the real-space distribution of $\kappa_{i}$ for a trefoil knot with $R:h=2:1$ (Figure 2b). Furthermore, $\kappa_{i}$ exhibits strict $2\pi/3$-periodicity versus $\theta_i$ in all models (Figures 2c and 2d), confirming that the magnitude of $\kappa_{i}$, directly reflecting the twisted geometry in 3D and the rotational symmetry of the knot with $C_3$, dictates the modulation of the SOC strength through geometric scaling.

\begin{figure*}[t]%
\centering
\includegraphics[width=0.85\textwidth]{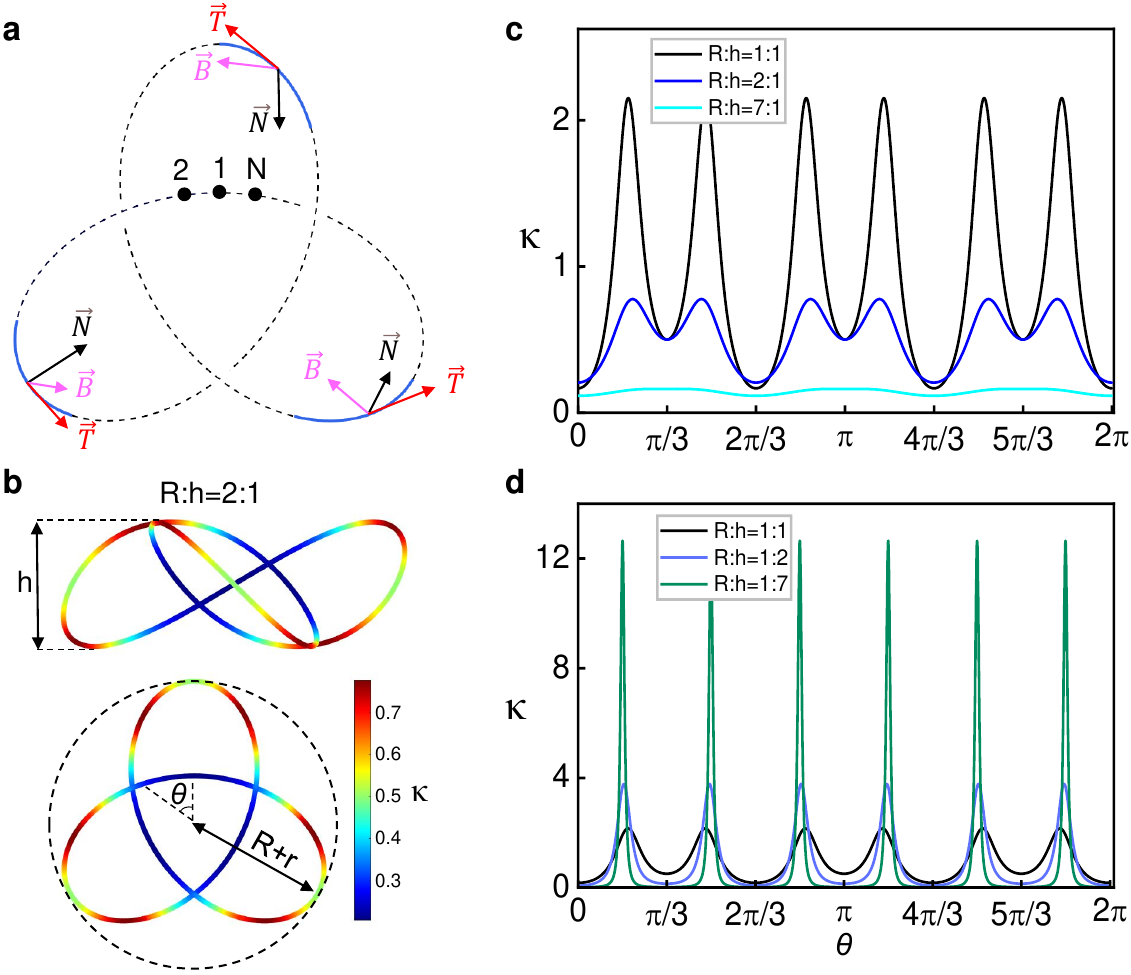}
\caption{\textbf{\textsf{Geometrical model and spatial curvature of trefoil knot molecules.}} \textsf{\textbf{a,} A trefoil knot model composed by $N$ lattices (top view), where the first and the last lattice sites are denoted, and the tangent ($\vec{T}_i$), normal ($\vec{N}_i$), and binormal vector ($\vec{B}_i$) are represented by red, black, magenta arrows, respectively. \textbf{b,} the lattice-dependent weigh distribution of spatial curvature $\kappa$, where four structural parameters $R$, $r$, $\theta$, and $h$ are denoted accordingly. \textbf{c} and \textbf{d,} The spatial curvature $\kappa$ versus the helix angle $\theta$ in the six representative trefoil knot models, with the specific structures drawn in SI, Figure S1.}}
\end{figure*}

For this trefoil knot model, given that $\mathcal{H}_{\mathrm{geo}}$ is determined, the overall tight-binding Hamiltonian can be formulated as $\mathcal{H}=\mathcal{H}_{\mathrm{mol}}+\mathcal{H}_{\mathrm{geo}}+\mathcal{H}_{\mathrm{el}}+\mathcal{H}_{\mathrm{d}}$, where $\mathcal{H}_\mathrm{mol}=\hat{T}+\hat{V}=\frac{\vec{p}^2}{2m}+V(x,y,z)$. Then, we discretize the Hamiltonian $\mathcal{H}_{\mathrm{mol}}$ and $\mathcal{H}_{\mathrm{geo}}$ within the Wannier representation, and the discretization form of the former is given by
\begin{equation}
\mathcal{H}_\mathrm{mol}=\sum^{N}_{n=1}\varepsilon_{n}c^{\dag}_{n}c_{n}+\sum^{N}_{n=1}(t_{n}c^{\dag}_{n}c_{n+1}+\mathrm{H}.\mathrm{c}.)
\end{equation}
where $c^{\dag}_{n}=(c^{\dag}_{n,\uparrow},c^{\dag}_{n,\downarrow})$ and $\varepsilon_{n}=\langle{\Psi_{n}}|\mathcal{H}_\mathrm{mol}|{\Psi_{n}}\rangle$ represent the creation operator and the onsite energy located at the $n$th lattice site, $t_{n}=\langle{\Psi_{n}}|\mathcal{H}_\mathrm{mol}|{\Psi_{n+1}}\rangle$ signifies the nearest-neighbor hopping integral between the $n$th and ($n$+1)th lattice sites and can be regarded as a constant. Here, ${\Psi_{n}}$ is the electron wave function at the $n$th lattice site. Similarly, $\mathcal{H}_\mathrm{geo}$ can be discretized as 
\begin{equation}
\mathcal{H}_\mathrm{geo}=
-i\frac{\lambda_{\mathrm{soc}}}{2}\sum^{N}_{n=1}c^{\dag}_{n}(\kappa_{n}+\kappa_{n+1})(\sigma_{n}+\sigma_{n+1})c_{n+1}+\mathrm{H}.\mathrm{c.}
\end{equation}
where $\lambda_{\mathrm{soc}}=\frac{\hbar^2}{16mS}$ with $S$ being the arc length between two adjacent lattices (SI, Section A). Within the framework of $(\vec{T}_i,\vec{N}_i,\vec{B}_i)$ and for each lattice, $\vec{B}_i$ is given by $\vec{B}_i=\vec{T}_i\times\vec{N}_i$. This unit vector has three components, namely $\vec{B}_i=(B_{ix}, B_{iy}, B_{iz})$, and then the Pauli matrix has the relation $\sigma_{i}=\sigma_x B_{ix}+\sigma_y B_{iy}+\sigma_z B_{iz}$, where $(\sigma_x, \sigma_y, \sigma_z)$ denotes the three dependent Pauli matrix. Unless otherwise specified, we adopt $\lambda_{\mathrm{soc}}$=0.737$t$ to perform all calculations. For example, $t$ = 0.3 eV resulting in $\lambda_{\mathrm{soc}}$ $\approx$ 0.221 eV, satisfies well the basic properties of electronic hopping and the geometric SOC established here in trefoil knot molecules.

To simulate the hopping of electrons between trefoil knot molecules, STM probes, and metal substrates, we adopt the coupling parameters $\Gamma_{\gamma}$, with $\gamma$ indicating the number of semi-infinite metal electrodes, to design a multi-terminal device model. Then, the Hamiltonian to describe the electron tunneling between the electrodes and the molecules can be described by $\mathcal{H}_\mathrm{el}=\sum_{n,\gamma} t'_n c^{\dag}_{n,\gamma}c_{n+1,\gamma}+\sum_{i,\gamma} \Gamma_{\gamma}c^{\dag}_{0,\gamma}c_{i}+\mathrm{H}.\mathrm{c}.$ ($n\geq 0$; $i$: lattice site in knot). $\mathcal{H}_\mathrm{d}$ represents the naturally occurring dephasing Hamiltonian in experiments. The dephasing process can arise from electron-phonon and electron-electron interactions. This inelastic scattering results in the loss of electron phase memory and can be simulated by connecting each position of the molecule to B\"{u}ttiker's virtual electrodes. Subsequently, under the boundary condition of zero net current on each virtual electrode, the spin-up conductance $G_{\uparrow}$ and spin-down conductance $G_{\downarrow}$ can be calculated by combining the Landauer-B\"{u}ttiker formula and non-equilibrium Green's functions, and then the spin polarization is defined as $P_s=\frac{G_{\uparrow}-G_{\downarrow}}{G_{\uparrow}+G_{\downarrow}}$ (see Methods).\\

\noindent
{\fontsize{11pt}{10pt}\selectfont \textbf{Fundamental conditions for knot-driven spin selectivity}}\\
\noindent
On the basis of the above Hamiltonian model, we establish the fundamental conditions for the emergence of spin polarization in conducting electron transport through trefoil knot molecules. To investigate this, we first construct a two-terminal device ($\gamma$=2) comprising a trefoil knot molecule ($N$=60 lattice sites) coupled to two non-magnetic metal electrodes (Figure 3a, upper panel). Theoretical calculations reveal that when the electrodes are coupled to the lattice sites $i$=1 and $j$=30, dividing the molecule into two perfectly symmetric halves, the system exhibits high conductivity but produces charge currents without spin polarization (Figure 3a, lower panel). This phenomenon arises from a pair of symmetric transport channels for tunneling electrons. Electrons traversing these channels in the curved space accumulate Berry phases, $e^{-i\Phi_{L}}$ and $e^{-i\Phi_{R}}$ ($L$, $R$ denote two parts divided by leads), with phase factors of equal magnitude but opposite sign. This phase opposition cancels quantum interference, thus suppressing spin polarization, a mechanism consistent with findings from our previous research on CISS in closed helical molecular rings {\color{blue}\cite{74,75,76}} (despite the fundamental difference in the type of SOC used in them). Furthermore, disconnecting the trefoil knot molecule at a specific point and positioning electrodes at both ends of this break (Figure 3a, right panel) yields numerical results indicating that while conductivity remains high (SI, Figure S2), significant spin polarization still does not emerge. This result confirms that a single-channel trefoil knot model cannot achieve spin polarization, aligning conclusively with previous conclusions derived from single-channel theoretical models of helix chains.

\begin{figure*}[t]%
\centering
\includegraphics[width=0.85\textwidth]{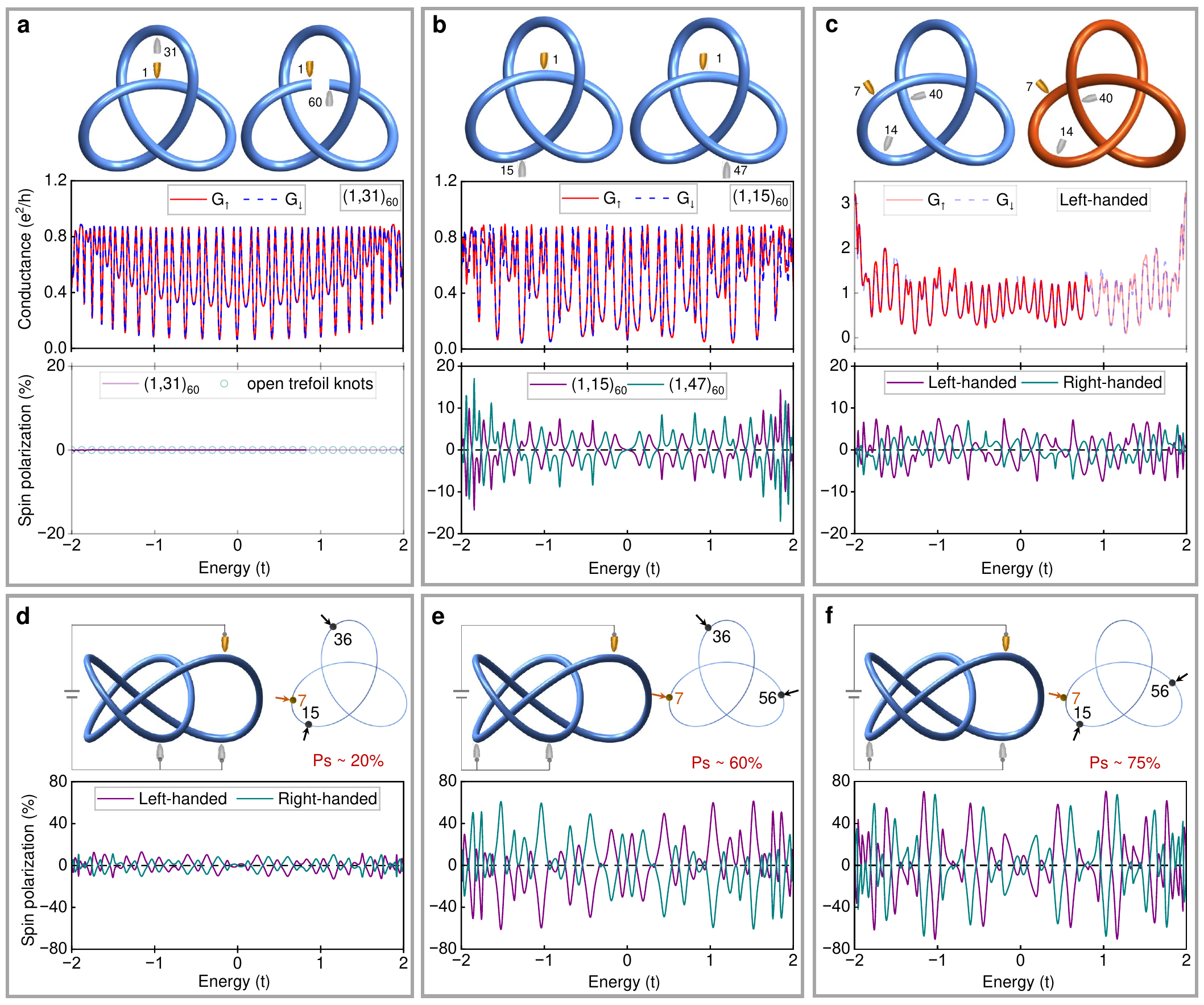}
\caption{\textbf{\textsf{Spin-dependent conductance (G$_\uparrow$ and G$_\downarrow$) and spin polarization $P_s$ in knot molecule-based devices.}} \textsf{\textbf{a,} Two two-terminal device configurations (top view): left device (1,31)$_{60}$ exhibits minor symmetry, while the right device features an open trefoil knot with ends coupled to electrodes; low panels show G$_\uparrow$, G$_\downarrow$ and $P_s$. \textbf{b,} Asymmetric two-terminal devices (1, 31)$_{60}$ and (1, 47)$_{60}$ with G$_\uparrow$, G$_\downarrow$ and $P_s$ plotted below. \textbf{c,} Three-terminal devices (7;14,40)$_{60}$ with opposite chirality and corresponding G$_\uparrow$, G$_\downarrow$ below. \textbf{d-f,} Three-terminal devices (7;15,36)$_{60}$, (7;36,56)$_{60}$ and (7;15,56)$_{60}$ with corresponding spin polarization.}}
\end{figure*}

When the lattice positions in the trefoil knot coupled by two electrodes break the local symmetry between the left and right transport paths, the Berry phase factors induced by spatial distortion no longer cancel each other out, leading to the emergence of spin polarization (Figure 3b). Specifically, in the configuration of the asymmetric two-terminal device (1,15)$_{60}$, the system exhibits both robust conductivity (SI, Figure S2) and prominent spin polarization of conducting electrons, reaching a peak magnitude of 20\%. In particular, the mirror-symmetric configuration (1,47)$_{60}$ exhibits completely opposite spin polarization (Figure 3b, lower part). This demonstrates that spatial inversion symmetry in the device design results in spin polarization of equal magnitude but opposite direction for these two configurations constructed on the same trefoil knot molecule. This finding not only aligns with the $C_3$ symmetry inherent to the trefoil knot, but also strongly validates the effectiveness and precision of the geometric SOC in explaining the underlying physics of TCISS. Consequently, the asymmetric multichannel transport mechanism is essential for spin-selective transport in trefoil knot molecules, with geometric SOC serving as a crucial factor in understanding the origin of spin polarization within this class of chiral molecules.

Considering the unique geometric configuration of the trefoil knot molecule and the prototypical device designed for experimental measurements of charge currents (Figure 1b), this molecule has three symmetrically arranged lowest lattice sites capable of coupling to a metal substrate, while any one of the three top lattices can be coupled to an STM probe. To model this, we develop a three-terminal device ($\gamma$=3) comprising a trefoil knot molecule with $N$=60: two of its lowest lattice sites are coupled to the metal substrate, and one of its highest lattice sites ($i$=7) is coupled to the STM probe. Specifically, three different coupling configurations for the lowest pair of sites connected to the substrate, corresponding to the structures shown in Figure 3d ($j$=15, 36), Figure 3e ($j$=36, 56), and Figure 3f ($j$=15, 56). Consequently, the spatial symmetry of the left and right transport channels formed by the bottom two electrodes and the top one progressively decreases across these three configurations. The calculation results reveal a sequential increase in spin polarization, from approximately 20\% to 75\%, directly demonstrating that asymmetric transport channels enhance spin selectivity.\\

\begin{figure*}[t]%
\centering
\includegraphics[width=0.9\textwidth]{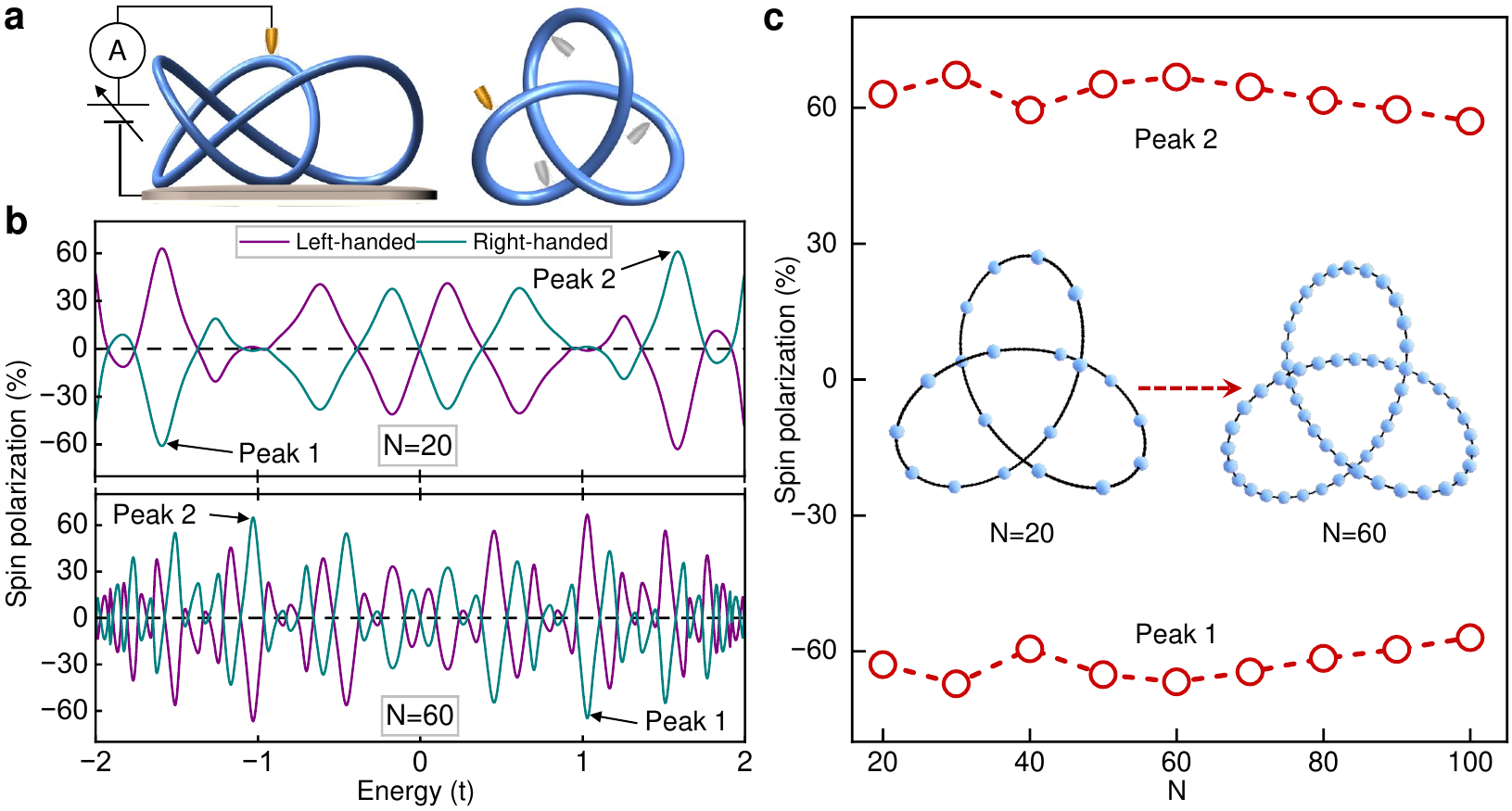}
\caption{\textbf{\textsf{Spin polarization of trefoil knot molecules versus the lattice density.}} \textbf{a,} A typical four-terminal device to simulate experiment measurements. \textbf{b,} Spin polarization of two typical trefoil knot models composed by $N$=20, and 60 lattices, where the curved length is fixed. \textbf{c,} Spin polarization of trefoil knot molecules versus the total lattice number $N$.}
\end{figure*}

\noindent
{\fontsize{11pt}{10pt}\selectfont \textbf{Robustness of high spin polarization in TCISS}}\\
\noindent
To gain a deeper understanding of the high spin polarization observed during electron transport in trefoil knot molecules, we construct a representative four-terminal device ($\gamma$=4) for experimental charge flow measurement {\color{blue}\cite{66,67}}: the three lowest lattice sites are coupled to a metal substrate, while the highest lattice site is coupled to an STM probe (Figure 4a). First, selecting a trefoil knot with $N$=20 (coupling sites $i$=3, $j$=6, 14, 20), the conductances $G_{\uparrow}$ and $G_\downarrow$ are calculated (Figure 4b, top panel). The numerical results reveal a significant spin polarization that reaches 60\%. Switching the molecular chirality from left- to right-handed reverses the direction of spin polarization, confirming its intrinsic dependence on the trefoil knot's topological chirality. Furthermore, this four-terminal configuration exhibits high conductivity, particularly near the Fermi level, where finite conductance indicates metallic transport behavior (SI, Figure S3). This is completely different from traditional chiral materials such as DNA and helicene, which show extremely low conductance because of significant band gaps. The observed metallic conductivity aligns with experiments that confirm an enhancement in the conductivity by nearly two orders of magnitude over conventional chiral materials. Additional calculations for $N$=60 lattice sites (constant molecular length) show that the maximum spin polarization remains $\sim$60\% (Figure 4b, bottom). In particular, this peak polarization ($\sim$60\%) is slightly lower than in the comparable three-terminal model ($\sim$75\%) with identical lattice count. This reduction arises because the fourth electrode improves the overall symmetry of the transport paths (consistent with $C_3$ rotational symmetry), partially counteracting the curvature-induced Berry phase factor and thereby suppressing spin polarization.

\begin{figure*}[t]%
\centering
\includegraphics[width=0.9\textwidth]{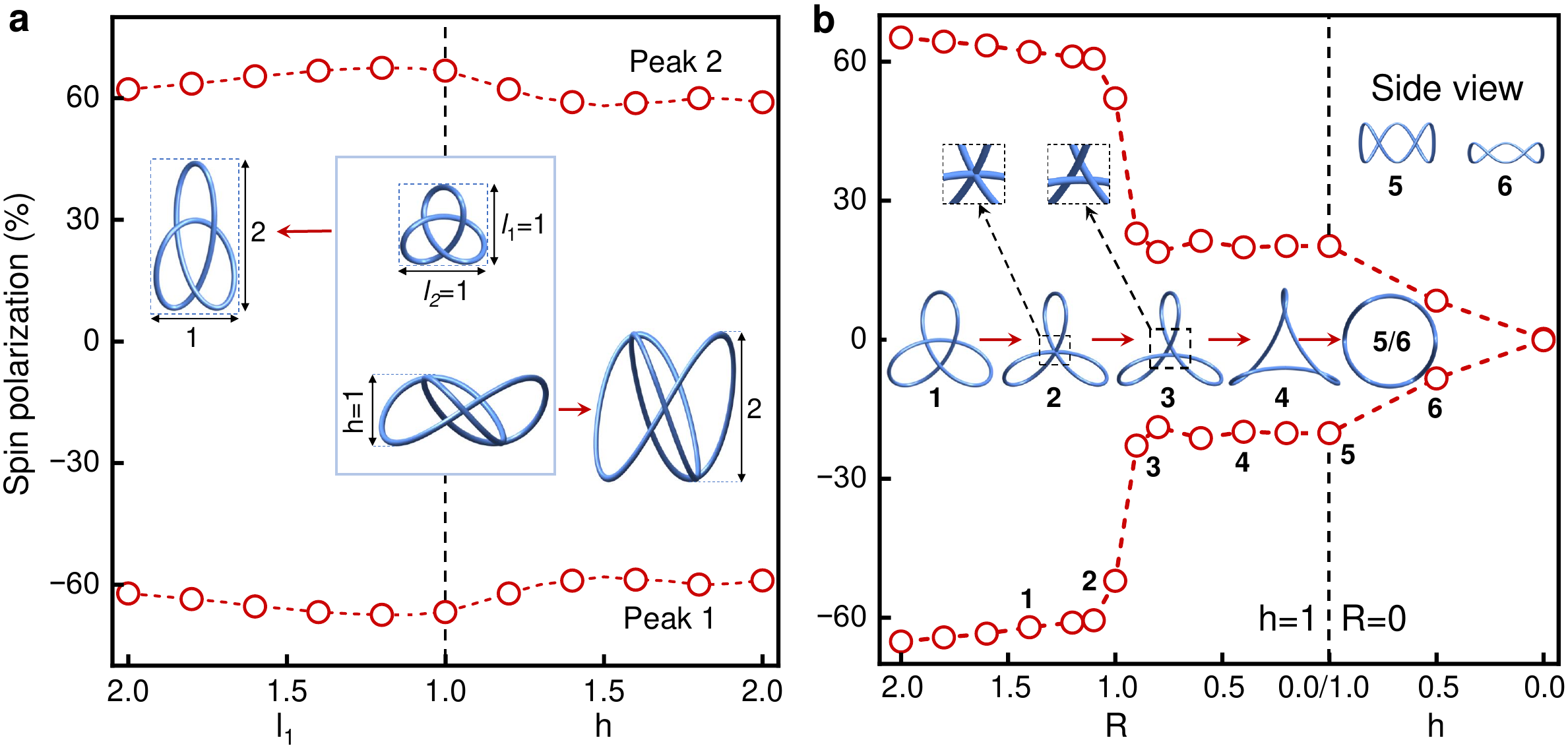}
\caption{\textbf{\textsf{Structural regulation of spin polarization in trefoil knot molecules.}} \textbf{a,} Strain effects: Left - Lateral strain  varies $l_1$ (1.0$\rightarrow$2.0, top view) while $l_2$ and $h$ is fixed; Right - Vertical strain increases height $h$ (1.0$\rightarrow$2.0, side view) with fixed $l_1$ and $l_2$. \textbf{b,} Chirality phase transition: Left - reducing structural parameter $R$ (2.0$\rightarrow$1.0) collapses the hollow triangle to a crossing point and reopens as a trivial ring with further reducing $R$ (1.0$\rightarrow$0.0, top view); Right - further reducing the ring height $h$ (1.0$\rightarrow$0.0). All systems share identical lattice number ($N$=60 in \textbf{a} and $N$=50 in \textbf{b}).}
\end{figure*}

Notably, in this four-terminal device based on the trefoil knot, spin polarization remains stable at approximately 60\% as the total lattice density increases ($N$=20 to 100) while maintaining constant molecular length (Figure 4c; see Video S1). These numerical results demonstrate that increasing the number of lattices does not effectively decrease spin polarization. Instead, polarization depends primarily on the molecule's intrinsic geometric configuration and coupling parameters, specifically, the number and position of coupled electrodes, rather than on the total lattice count. This behavior sharply contrasts with the conventional CISS effect in helical chains, where spin polarization typically exhibits significant variation, often increasing positively with either the number of lattices per unit cell or the overall chain length. Collectively, these observations reveal that the TCISS effect originates from the intrinsic topological property of the molecule: The stable knotted geometry ensures topological invariance, thereby maintaining consistent spin polarization in TCISS.

To verify the stability of high spin polarization in the TCISS effect, we use a four-terminal device ($\gamma$=4) composed of an $N$=60 trefoil knot molecule ($l_1$=$l_2$=1.0, $h$=1.0; central structures in Figure 5) for strain manipulation. Based on its 3D helical structure, two strain strategies are implemented: (i) lateral strain that maintains constant height ($h$=1.0) while compressing one side asymmetrically ($l_1$=1.0$\rightarrow$2.0, $l_2$=1.0; Figure 5a, left), and (ii) vertical strain that maintains longitudinal and lateral sizes while stretching height ($h$=1.0$\rightarrow$2.0; Figure 5a, right). Calculations reveal that lateral deformation does not induce significant changes in either conductivity (SI, Figure S4) or spin polarization ($\sim$60\%). Similarly, vertical deformation does not produce significant polarization changes relative to the unstrained knot (Figure 5a, right curves). This persistence of high spin polarization under structural deformation robustly demonstrates the structural stability of the TCISS effect, attributed to the symmetry-protected knot topology. These findings provide critical theoretical insight for experimental observations: trefoil knots maintain ultra-stable spin polarization at high temperatures because lattice vibration-induced deformations cannot disrupt topologically protected spin-selective channels. This fundamental distinction between the TCISS and the conventional CISS highlights the unique size-independent robustness and performance advantages of trefoil knots in spintronics applications.\\

\noindent
{\fontsize{11pt}{10pt}\selectfont \textbf{Topological phase transition of TCISS}}\\
\noindent
To verify the critical role of knot topology in high spin polarization and robustness of the TCISS effect, we modulated the structural parameter $R$ in the trefoil knot with $N$=50. Variations in $R$ suppress or eliminate knot topology while profoundly impacting spin polarization in charge transport. As $R$ decreases from 2.0 to 1.1, the central triangular area shrinks but retains the knot topology (Structure 1, Figure 5b; Video S2), with the spin polarization stable at $\sim$60\%. At the critical value $R$=1.0, the central region collapses to a crossing point (Structure 2, Figure 5b), causing a significant reduction in polarization despite finite values. Further reduction ($R$=0.9, Structure 3, Figure 5b) eliminates the topological knot: The helical ring unfolds into a trivial knot, transformable to a single annular pore, yet retains fan-like structural chirality (Structures 3-4, Figure 5b; see Video S2). This deformation induces a topological phase transition where topological chirality yields to structural chirality. Crucially, post-transition spin polarization falls to $\sim$20\%, indicating degradation of TCISS to conventional CISS with substantially reduced conductivity (SI, Figure S5). This sequential transformation unambiguously demonstrates that the stability of $>$60\% spin polarization originates from topological protection by the nontrivial knot.

Crucially, progressive reduction of $R$ transforms the trefoil knot into a fan-shaped configuration. At $R$=0, the system extends completely into a circular vertical ring (Structures 5, 6 from top view, Figure 5b; Video S2). Despite this deformation, asymmetric transport paths with non-zero curvature preserve finite spin polarization in conducting electrons. In contract, the reducing height parameter $h$ flattens the cyclic molecule (Structures 5-6 from side view, Figure 5b). At $h$=0, the structure degenerates to a circular planar ring with a uniform $\kappa$, causing the spin polarization to decay asymptotically to zero. This sharp contrast demonstrates that the structurally chiral CISS exhibits extreme geometric sensitivity, which is fundamentally different from the TCISS revealed in this study. Notably, the geometric SOC framework employed here provides a unified physical basis for both conventional CISS and TCISS uncovered here.\\


\noindent
{\fontsize{11pt}{10pt}\selectfont \textbf{Discussion}}\\
\noindent
This study pioneers a theoretical framework that elucidates the microscopic origin of ultra-high spin polarization in trefoil knot molecules, termed the TCISS effect. The mechanism comprises two essential components: asymmetric multiple transport channels and geometric SOC. The latter originates from quantum geometric phase accumulation in curved configuration spaces, which is fundamentally different from the conventional SOC invoked in the explanations on traditional CISS. We develop multiterminal device models that incorporate electronic hopping and geometric SOC through tight-binding Hamiltonian models. Theoretical calculations demonstrate that electrode coupling that breaks spatial symmetry while preserving molecular knot connectivity generates significant spin polarization. Realistic multiterminal simulations, considering single-particle interactions without electron correlation enhancements, reveal charge currents exhibiting $>$60\% spin polarization with a direction exclusively determined by topological chirality, alongside high conductivity. These predictions quantitatively explain experimental observations, establishing a foundational microscopic mechanism for spin selectivity in topological chiral materials.

Theoretical calculations further demonstrate that when maintaining constant molecular length while varying lattice site density, both high spin polarization ($>$60\%) and high conductivity remain stable. Under lateral/vertical deformation that reaches the 200\% amplitude, these properties exhibit exceptional structural robustness. However, suppressing topological knots, triggering their degeneration into trivial knots with a concomitant transition from topological chirality to structural chirality, causes a drastic reduction in spin polarization. This unequivocally confirms that the TCISS effect in trefoil knots enjoys topological protection and exhibits superior robustness. This characteristic directly explains the persistence of ultrahigh spin polarization observed experimentally, even in high-temperature regimes.

Our theoretical model yields 60\% spin polarization for trefoil knot molecules, significantly below experimentally reported values that approach 90\% {\color{blue}\cite{66,67}}. This discrepancy arises crucially from simplified model conditions: a single-molecule device in which only one top lattice site couples to the STM probe. However, in actual experiments, the STM probe interacts with multiple lattice sites or knot molecules, a configuration that amplifies spin polarization. Furthermore, experimental samples comprise 2-20 nm molecular thin films where spin-polarized currents traverse multiple molecular layers, further increasing polarization. Incorporating these experimental factors and intrinsic enhancement mechanisms (e.g., electron-electron correlations and electron-phonon coupling) would align theoretical predictions with the observed 90\% polarization. Thus, when adjusting for experimental scalability and intrinsic effects, this ultrahigh spin polarization demonstrates both theoretical predictability and physical consistency.

Beyond the above innovations, our theory provides the first successful elucidation of CISS effect variations during phase transitions between topological chirality and structural chirality. Although established for the TCISS in trefoil knots, this theoretical framework can be extended to other chiral knot molecules, such as the cinquefoil (5$_1$) knot {\color{blue}\cite{77,78}}, creating a new chiral spintronics platform based on knotting matter. More importantly, this approach offers a unified perspective for uncovering CISS physical mechanisms across diverse chiral materials, representing a critical advancement toward a universal theory of CISS phenomena.\\

\noindent
{\fontsize{11pt}{10pt}\selectfont \textbf{Methods}}\\
\noindent
\textbf{Geometric SOC calculation.} The detailed derivations of the geometric SOC within trefoil knot molecules are provided in Supplementary Information, Section A.\\

\noindent
\textbf{Spin-polarized conductance and spin polarization.} To introduce the theoretical method to calculate conductance, we take a two-terminal device ($\gamma$=2) as an example. By introducing the spin degree of freedom and using the parameters $\nu=L,R$ to denote the two metal electrodes, that is, the left and right leads, while employing the parameter $n$ to signify semi-infinite virtual electrodes at each lattice site, we then construct the trefoil knot molecule-based device to simulate experimental measurements on charge currents. Within the framework of the Landauer-Büttiker formula, the charge current $I_\nu$ flowing in the electrode $\nu$ and the current $I_n$ flowing in the virtual electrode $n$ can be computed below
\begin{equation}
\begin{aligned}
J_{\nu}=\frac{e}{\hbar}\sum_{\alpha,\sigma}\int^{\infty}_{-\infty}T^{\sigma}_{L,\alpha}[f^{\nu}(E)-f^{\alpha}(E)]dE,\\
J_{n}=\frac{e}{\hbar}\sum_{\alpha,\sigma}\int^{\infty}_{-\infty}T^{\sigma}_{n,\alpha}[f^{n}(E)-f^{\alpha}(E)]dE,
\end{aligned}
\end{equation}
The Fermi-Dirac distribution within an electrode is expressed in terms of the inverse temperature $\beta=(k_B T_{\mu})^{-1}$ and the chemical potential $\mu$, denoted as $f_{\mu}(E)=[e^{\beta_{\nu}(E-\mu_{\nu})}+1]^{-1}$. Consequently, to make sure the currents $I_{n}=0$ in each probe constitute a set of non-linear equations that are interlinked with the local chemical potentials.
\begin{equation}
\sum_{\alpha,\sigma}\int^{\infty}_{-\infty}T^{\sigma}_{n,\alpha}f^{n}(E)dE=\sum_{\alpha,\sigma}\int^{\infty}_{-\infty}T^{\sigma}_{n,n'}f^{n'}(E)dE+\sum_{\alpha,\sigma}\int^{\infty}_{-\infty}T^{\sigma}_{n,\nu}f^{\nu}(E)dE
\end{equation}
To simplify the current calculations, we restrict the computation of $N$ local currents within the scope of linear response for this matrix equation. Furthermore, under low-voltage conditions, we can employ the Taylor series to expand the Fermi-Dirac distribution function to the first nontrivial order, thereby facilitating the approximation of electronic transport characteristics. Meanwhile, the Fermi-Dirac distribution function can be approximately expanded as
\begin{equation}
f^{\alpha}(E,\mu_\alpha)=f^{eq}(E,F_F)-\frac{\partial f^{eq}(E,E_F)}{\partial{E}}(\mu_{\alpha}-E_F),  (\mu_{eq}=E_F),
\end{equation}
With the Fermi energy set to zero, the above equations can be simplified as follows
\begin{equation}
\mu_{n}\sum_{\alpha,\sigma}\int^{\infty}_{-\infty}T^{\sigma}_{n,\alpha}(-\frac{\partial f^{eq}(E,E_F)}{\partial{E}})-\sum^N_{n',\sigma}\int^{\infty}_{-\infty}T^{\sigma}_{n,n'}(-\frac{\partial f^{eq}(E,E_F)}{\partial{E}})\\
=\int^{\infty}_{-\infty}(-\frac{\partial f^{eq}(E,E_F)}{\partial{E}})[T^{\sigma}_{n,L}(E)\mu_{L}+T^{\sigma}_{n,R}(E)\mu_{R}].
\end{equation}
In this way, the matrix equation comprising $n$ equations can be reformulated as an equation $\hat{M}\cdot{\mu}=\nu$, and the solutions to this matrix equation can be obtained by inversion of the matrix. Furthermore, the solution of this equation will allow us to determine the specific form of the chemical potential associated with the probe coupled to each lattice site. The matrix $\hat{M}$, in this context, can be expressed as
\begin{gather}
{\hat{M}=
\begin{bmatrix}
    \sum_{\alpha\neq{1},\nu,\sigma}\int^{\infty}_{-\infty}dET^{\sigma}_{1,\alpha}(E)(-\frac{\partial f^{\alpha}(E,E_F)}{\partial{E}})&   \ldots&   -\sum_{\sigma}\int^{\infty}_{-\infty}dET^{\sigma}_{1,n}(E)(-\frac{\partial f^{\alpha}(E,E_F)}{\partial{E}})&    \\
    -\sum_{\sigma}\int^{\infty}_{-\infty}dET^{\sigma}_{2,1}(E)(-\frac{\partial f^{\alpha}(E,E_F)}{\partial{E}})&   \ddots&    -\sum_{\sigma}\int^{\infty}_{-\infty}dET^{\sigma}_{2,n}(E)(-\frac{\partial f^{\alpha}(E,E_F)}{\partial{E}})&    \\
    \ldots&   \ldots&   \sum_{\alpha\neq{n},\nu,\sigma}\int^{\infty}_{-\infty}dET^{\sigma}_{n,\alpha}(E)(-\frac{\partial f^{\alpha}(E,E_F)}{\partial{E}})&   \\
\end{bmatrix}.}
\end{gather}
By solving the matrix equation mentioned above, we can determine the chemical potential of the virtual probe and subsequently calculate the spin-dependent charge currents flowing from the metal electrodes as below
\begin{equation}
J_{L,\sigma}=\frac{e^2}{\hbar}\sum_{\alpha\neq{1},\nu}T^{\sigma}_{L,n}(E)[-\frac{\partial f^{\alpha}(E,E_F)}{\partial{E}}](\mu_{L}-\mu_{R}).
\end{equation}
Within the linear response regime, the conductivity is simply given by $G_{\sigma}=\frac{J_{L,\sigma}}{\Delta{V}}$, where $\Delta{V}=\mu_{R}-\mu_{L}$. Furthermore, under low-temperature conditions, the Fermi distribution function reduces to the Dirac delta function, and the spin conductivity can be represented as follows
\begin{equation}
G_{\sigma}=\frac{e^2}{\hbar}[T^{\sigma}_{L,R}(E_F)+\sum_{n,\sigma}T^{\sigma}_{L,n}(\mu_{L}-\mu_{n})/\Delta{\mu}],
\end{equation}
Thus, we will derive the specific expression for spin polarization $P_s$ as
\begin{equation}
P_s=\frac{G_{\uparrow}-G_{\downarrow}}{G_{\uparrow}+G_{\downarrow}}.
\end{equation}\\

\noindent
{\fontsize{11pt}{10pt}\selectfont \textbf{Acknowledgments}}\\
\noindent
This work is supported by the National Natural Science Foundation of China with Grant Nos. 11774104 and U1832209, and partly by the National Key R\&D Program
of China (2021YFC2202300).\\

\noindent
{\fontsize{11pt}{10pt}\selectfont \textbf{Authors contributions}}\\
\noindent
H.F. conceived and designed the project. X.S. and K.Z. performed the theoretical calculations. H.F. and X.S. wrote the original draft. H.F. improved the manuscript and determined its final version with contributions from all authors.\\ 

\noindent
{\fontsize{11pt}{10pt}\selectfont \textbf{Competing interests}}\\
\noindent
The authors declare no competing interests.\\

\noindent
{\fontsize{11pt}{10pt}\selectfont \textbf{Additional information}}\\
\noindent
\textbf{Supplementary information} The online version contains supplementary material available at ...\\

\noindent
\textbf{Correspondence and requests for materials} should be addressed to H.F.


\end{document}